\documentclass[prd, onecolumn, floatfix, nofootinbib, notitlepage, showkeys, showpacs]{revtex4-1}

\usepackage[mathscr]{euscript}
\usepackage{amsmath}
\usepackage{graphicx}
\usepackage{dcolumn}
\usepackage{bm}
\usepackage{epsfig}
\usepackage{amssymb,latexsym,mathrsfs}
\usepackage{graphicx}
\usepackage{color}
\usepackage{hyperref}
\usepackage{float}
\usepackage{diagbox}

\usepackage{tikz}
\usepackage[compat=1.1.0]{tikz-feynman}

\usepackage{amsmath}

\hypersetup{
    colorlinks=true,
    linkcolor=red,
    citecolor=blue,
}

\usepackage{amsmath}
\usepackage{amssymb}
\usepackage{subfigure}
\usepackage{hyperref}
\usepackage{url}
\usepackage{xcolor}
\usepackage{color}
\definecolor{amaranth}{rgb}{0.9, 0.17, 0.31}
\definecolor{purple(munsell)}{rgb}{0.62, 0.0, 0.77}
\definecolor{americanrose}{rgb}{1.0, 0.01, 0.24}
\definecolor{palatinateblue}{rgb}{0.15, 0.23, 0.89}
\definecolor{royalblue(web)}{rgb}{0.25, 0.41, 0.88}
\definecolor{hanpurple}{rgb}{0.32, 0.09, 0.98}
\definecolor{beaublue}{rgb}{0.74, 0.83, 0.9}
\definecolor{carminered}{rgb}{1.0, 0.0, 0.22}
\definecolor{brightpink}{rgb}{1.0, 0.0, 0.5}
\definecolor{vividviolet}{rgb}{0.62, 0.0, 1.0}

\definecolor{electron}{rgb}{1.0, 0.67, 0.22}
\hypersetup{ linktoc=all,
    colorlinks, linkcolor={palatinateblue},
    citecolor={brightpink}, urlcolor={amaranth}}

\newcommand{\changeurlcolor}[1]{\hypersetup{urlcolor=#1}}

\newcommand{\be}{\begin{equation}}
\newcommand{\ee}{\end{equation}}
\newcommand{\bs}{\begin{split}} 
\newcommand{\bea}{\begin{eqnarray}}
\newcommand{\eea}{\end{eqnarray}}

\newcommand{\bes}{\begin{subequations}}
\newcommand{\ees}{\end{subequations}}

\renewcommand{\d}[1]{\ensuremath{\operatorname{d}\!{#1}}}

\renewcommand{\d}[1]{\ensuremath{\operatorname{d}\!{#1}}}

\newcommand{\bo}{\raise-1mm\hbox{\Large$\Box$}}

\begin{document}


\title{Spacetime continuity and quantum information loss}

\author{Michael R. R. Good${}^{1}$}
\email{michael.good@nu.edu.kz}

\affiliation{
${}^1$ Nazarbayev University, Astana 010000, Kazakhstan
}


\begin{abstract} 
Continuity across the shock wave of two regions in the metric during the formation of a black hole can be relaxed in order to achieve information preservation. A Planck scale sized spacetime discontinuity leads to unitarity (a constant asymptotic entanglement entropy) by restricting the origin of coordinates (moving mirror) to be timelike.   Moreover, thermal equilibration occurs and total evaporation energy emitted is finite.
\end{abstract} 

\date{\today} 

\maketitle



\section{Introduction}



In this note, the role of continuity and information loss (via the tortoise coordinate, $r^*$), in understanding the crux of the phenomena of particle creation from black holes \cite{Hawking:1974sw} is explored.  In particular the relationship to the simplified model of the moving mirror \cite{Davies:1976hi,Davies:1977yv,Fulling:2018lez}, is investigated because it has identical Bogolubov coefficients \cite{Good:2016oey}.  

Uncompromising continuity is relaxed in favor of unitarity by a single additional parameter generalization of the $r^*$ coordinate. An understanding of the correspondence between the black hole and the moving mirror in this new context is initiated \cite{Good:2016atu, GTC, Myrzakul:2018bhy}. We prioritize information preservation, and find finite evaporation energy, thermal equilibrium, and analytical beta coefficients. Moreover, we determine and assess a left-over remnant (e.g. \cite{Wilczek:1993jn, remnants}).

Broadly motivating, delving into the ramifications of external effects on quantum fields have led to interesting results on a wide variety of phenomena all the way from, e.g. relativistic superfluidity \cite{super,oxford} to the creation of quantum vortexes in analogue spacetimes \cite{vortex,Liberati:2018uev}.  The fertile enterprise of QFT under external conditions (like the external condition of a rotating black hole spacetime) has resulted in simple and revealing relations, (e.g. \cite{springy} $2\pi T = g - k$, temperature of a Kerr black hole).



Specifically motivating, the physical effect of the black hole origin is that it amplifies quantum field fluctuations by reflecting virtual particles into real ones. There is however, a problem:  the black mirror \cite{Good:2016oey}, which is the aforementioned usual tortoise coordinate associated boundary condition, extracts energy indefinitely and does not preserve unitarity. 

We sketch a possible simple way out - `a timelike worldline for the origin' - which resolves these two problems (as probed in \cite{Good:2016atu}). It was demonstrated recently \cite{GTC,Myrzakul:2018bhy} that a timelike worldline compels the introduction of a second parameter, $\xi$, the asymptotic sub-light speed of the origin (in addition to mass, $M$), which generalizes the tortoise coordinate and characterizes the dynamics, evaporative energy, spectra, information, and continuity.

\section{Asymptotic Null Origin} 
\begin{figure}[h]
\centering 
\includegraphics[width=2.6in,height=3.0in,keepaspectratio]{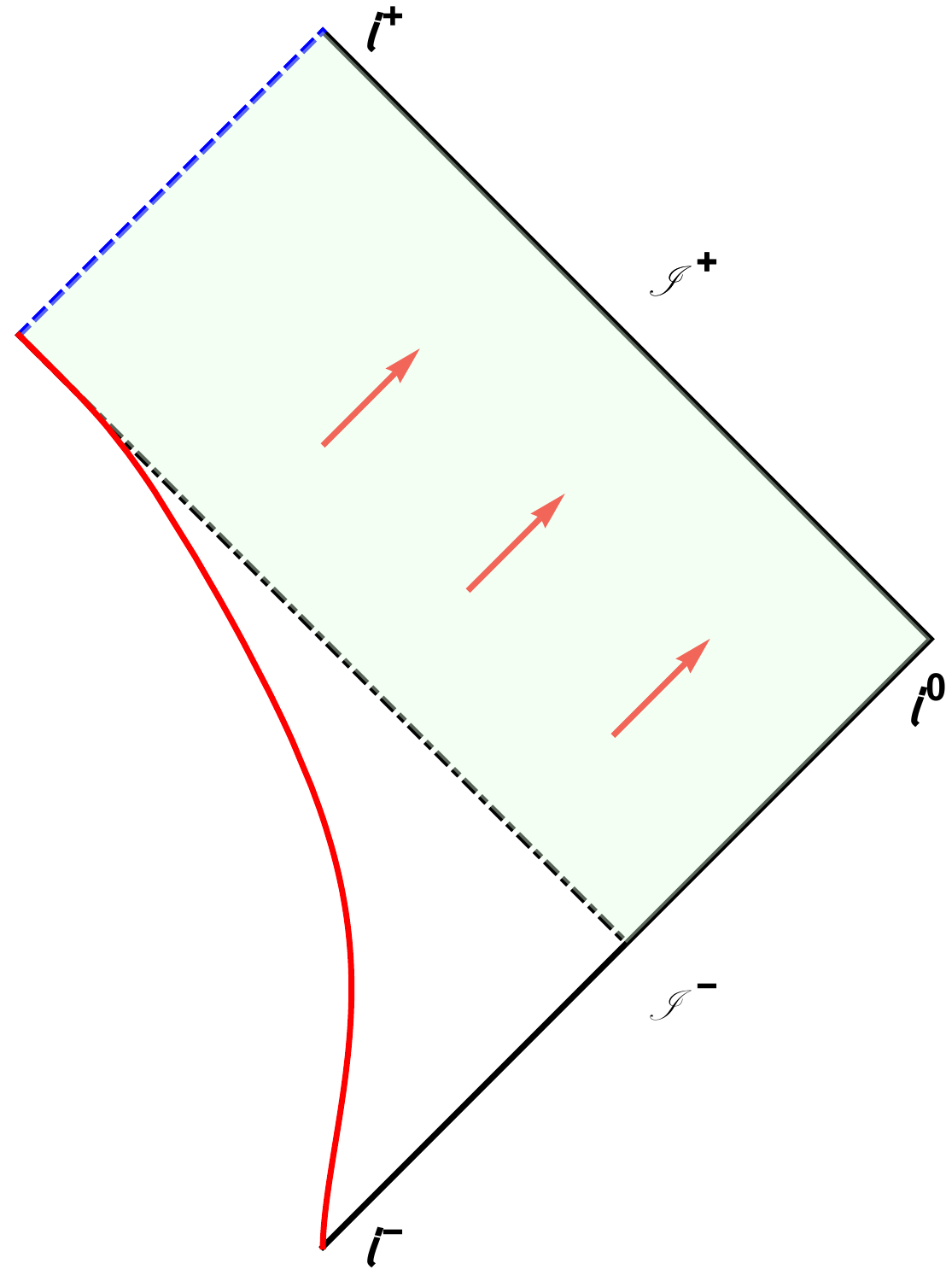} 
\;\;\;\;\;\;\;\;\;\;\;
\includegraphics[width=2.6in,height=3.0in,keepaspectratio]{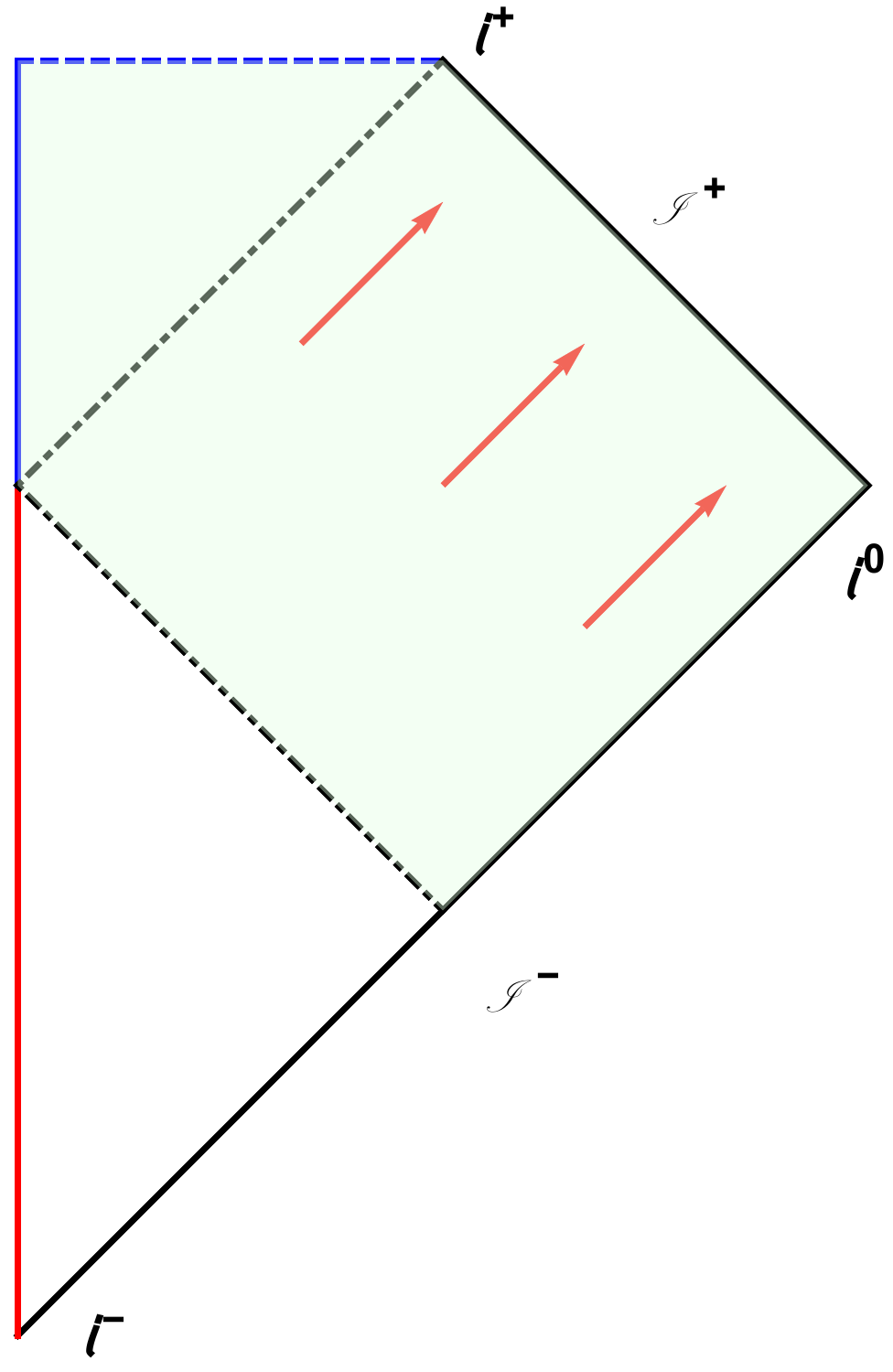} 
\caption{\textbf{Left}: The 1+1 dimensional Penrose diagram for the black hole collapse mirror.  The red line is the mirror, Eq.~(\ref{omex}), computed with $\kappa=2$ for illustration. The dotted black line is the horizon, $v_H= 0$.  The blue dashed line is the upper-half of the left future null-infinity Cauchy surface, $\mathscr{I}_L^+$ (unlabeled), that acts as a singularity analog. Notice the light green opaque region is where left-moving modes never reflect off the mirror and get `trapped', never to become right-movers.  The red arrows are Hawking radiation emitted by the mirror or they can be left-movers that reflected off the mirror into right-movers reaching $\mathscr{I}^+$. \textbf{Right}:  The usual 3+1 dimensional Schwarzschild causal structure captured in a Penrose diagram.  The blue dashed line is the space-like singularity while the blue solid line is $r=0$ where modes pass through but still hit the singularity.  The black dotted dashed line is the event horizon.  The red arrows are Hawking radiation or they can be left-movers that passed through $r=0$, `reflected' and become right-movers.  They will eventually escape to the observer at $\mathscr{I}^+$. Notice the light green opaque region where left-movers are lost into the singularity.  } 
\label{infolossBHC} 
\end{figure}

\begin{figure}[h]
\centering 
\includegraphics[width=2.6in,height=3.0in,keepaspectratio]{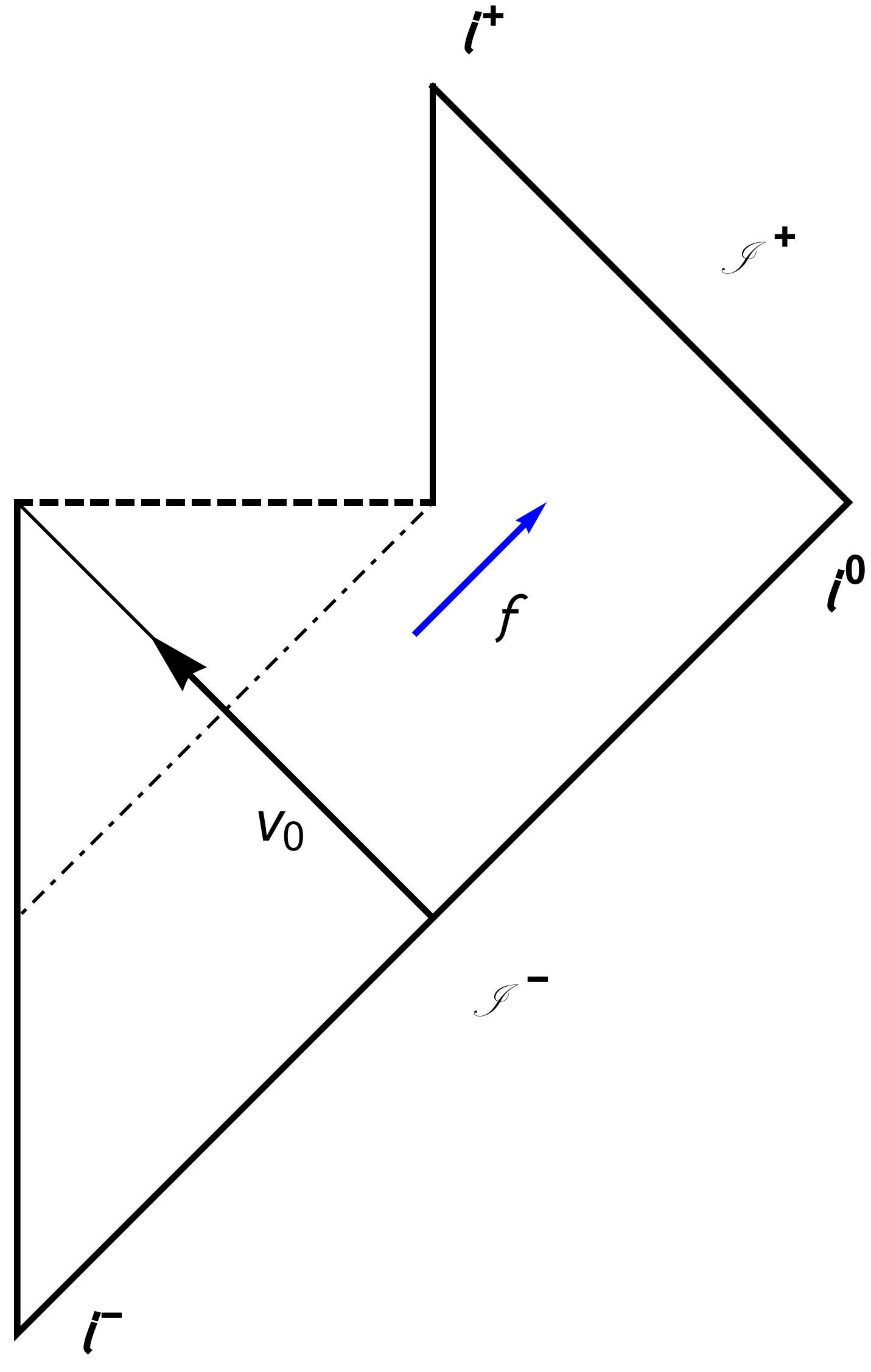} 
\includegraphics[width=3.0in,height=3.0in,keepaspectratio]{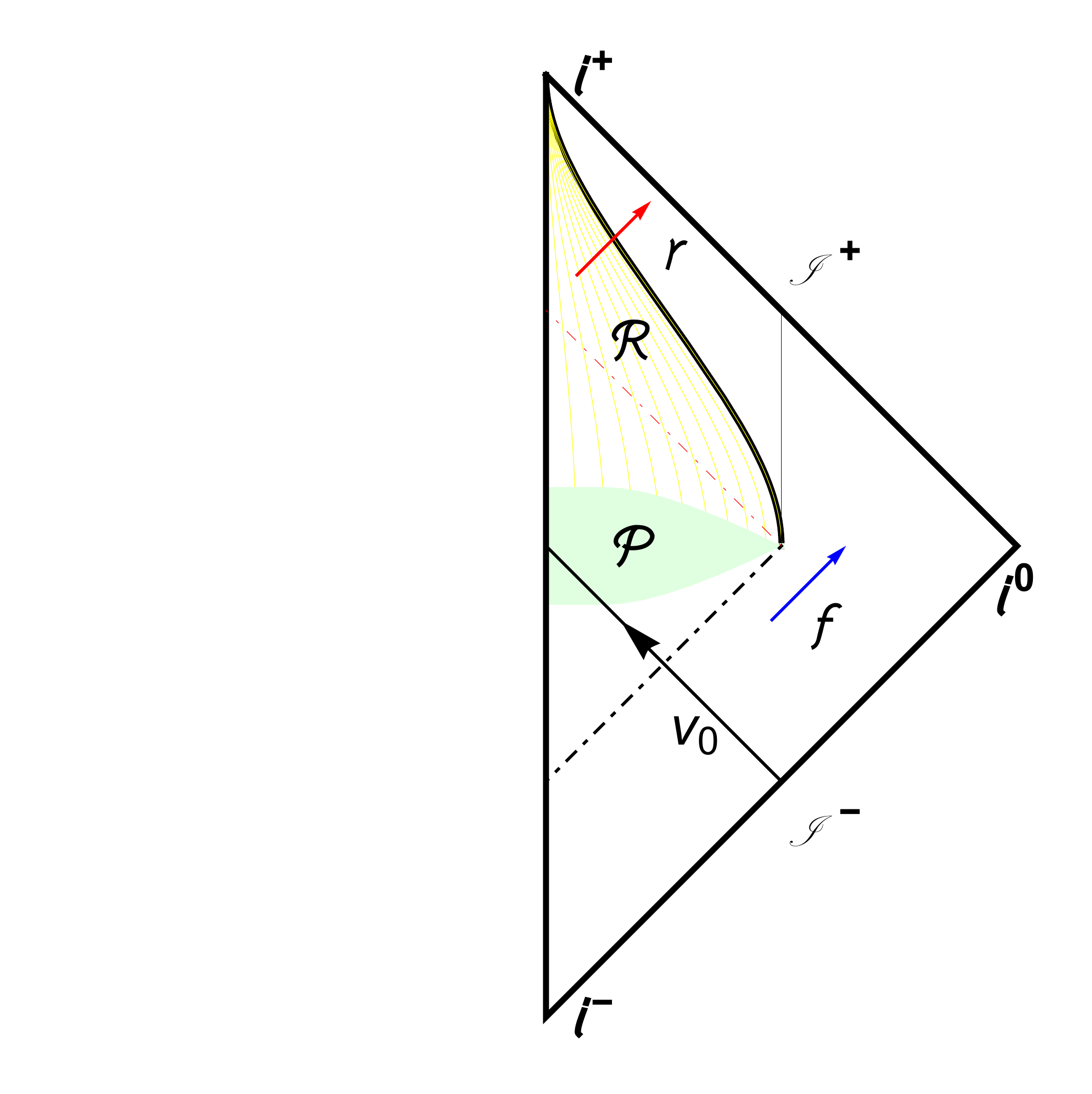} 
\caption{\textbf{Left}: The usual Penrose diagram for an evaporating Schwarzshild black hole with a singularity (thick dashed horizontal line) and information loss. The dotted dashed line is the event horizon. The outgoing blue arrow, $\mathit{f}$, is the Hawking flux and the ingoing black arrow, $v_0$, is the shock wave. \textbf{Right}: Penrose diagram for the 3+1 dimensional case where a black hole remnant remains with infinite lifetime, for a similar diagram see \cite{remnants}. Notice the redshifted outgoing modes labeled as $\mathit{r}$ with a red arrow, escape to null-infinity $\mathscr{I}^+$.  Even though Hawking evaporation stops, all field modes passing through the remnant undergo an eternal late-time  redshift after passing through and out of the remnant region $\mathcal{R}$.  The light green area labeled $\mathcal{P}$ is the Planckian region where quantum geometric effects are important (discontinuity) and has a non-zero, maximum, finite height centered at $r=0$.  All field modes can `reflect' off of $r=0$ and pass through the $\mathcal{R}$ region without loss of information. } 
\label{fig:penrose} 
\end{figure}

It is first appropriate to survey the situation \textit{with} information loss.  This is plotted as the usual Penrose diagram in Fig. (\ref{fig:penrose}).  To that end, the matching solution for the outside and inside of the black hole over the shock wave is derived, following the notation of Wilczek \cite{Wilczek:1993jn} and in the textbook of Fabbri-Navaro-Salas \cite{Fabbri:2005mw}. The Regge-Wheeler coordinate, $r^*$,
\begin{equation}
r^*\equiv r+2M \ln\left(\frac{r}{2M}-1\right),  \label{tortoise}
\end{equation}
sets the dynamics.  In the usual set up, the metric is matched on both sides of the shock wave, $v_0$, inside and out, respectively:
\begin{equation}
r(v_0, U)=r(v_0, u), \label{rmatch}
\end{equation}where 
\begin{equation}
r(v_0, U)=\frac{v_0-U}{2},\quad \quad 
\text{and} \quad\quad
r^*(v_0, u)=\frac{v_0-u}{2}.\label{rnull}
\end{equation} 
Using Eq.~(\ref{tortoise}), Eq.~(\ref{rmatch}) and Eq.~(\ref{rnull}) to solve for the trajectory of the origin, $u$, in null coordinates gives:
\begin{equation}
u=U-4M \ln|\kappa U|.  \label{m1}
\end{equation}
Here we define the surface gravity, $\kappa \equiv 1/4M$. As $u\to +\infty$, the formation of a event horizon occurs at $v_H\equiv v_0-4M$, where for simplicity and without loss of physical generality we set $v_H=0$ (shock wave at $v_0=4M$). The result, Eq.~(\ref{m1}), is the matching solution \cite{Fabbri:2005mw,Wilczek:1993jn,GTC} required for the Schwarzschild geometry (exterior) to the Minkowski geometry (interior) with a strict event horizon.  The regularity condition (the field must be zero at $r=0$) restricts the behavior of the modes in the interior such that $U\leftrightarrow v$. Converting to Cartesian coordinate $x$ by the definitions $ u \equiv t(x) - x$ and $v \equiv t(x) + x$ into Eq.~(\ref{m1}), and solving for $t(x)$ gives the equation of motion of the black mirror\footnote{The black mirror, which is not early-time thermal (like the Carlitz-Willey trajectory \cite{Good:2012cp,Carlitz:1986nh}), is often called Omex after the transcendental \textbf{Om}ega, $\Omega e^{\Omega} = 1$, and \textbf{ex}ponential \cite{Good:2016oey}. It is also called the Black Hole Collapse (BHC) trajectory \cite{Cong:2018vqx}}:
\be t(x) = - x - 4M e^{x/2M},\label{omex}\ee
which is studied\footnote{Likewise see MG14 proceedings\cite{Good:2015jwa,Anderson:2015iga} and the 2nd LeCosPA Symposium \cite{Good:2016bsq}.} in \cite{Good:2016oey}.  
 This is plotted as mirror in Fig. (\ref{fig:penrose_mirror}). The coordinates span $0<r<\infty$ and $-\infty < x < \infty$. The transcendentally invertible trajectories of Eq.~(\ref{m1}) and Eq.~(\ref{omex}) characterize a one-parameter system (the mass $M$) and determine the late-time (thermo)dynamics which results in information loss. See Fig.~(\ref{infolossBHC}) for a visual of how modes traveling through the horizon are `lost'.

\section{Asymptotic Timelike Origin} 
\begin{figure}[h]
\centering 
\includegraphics[width=2.6in]{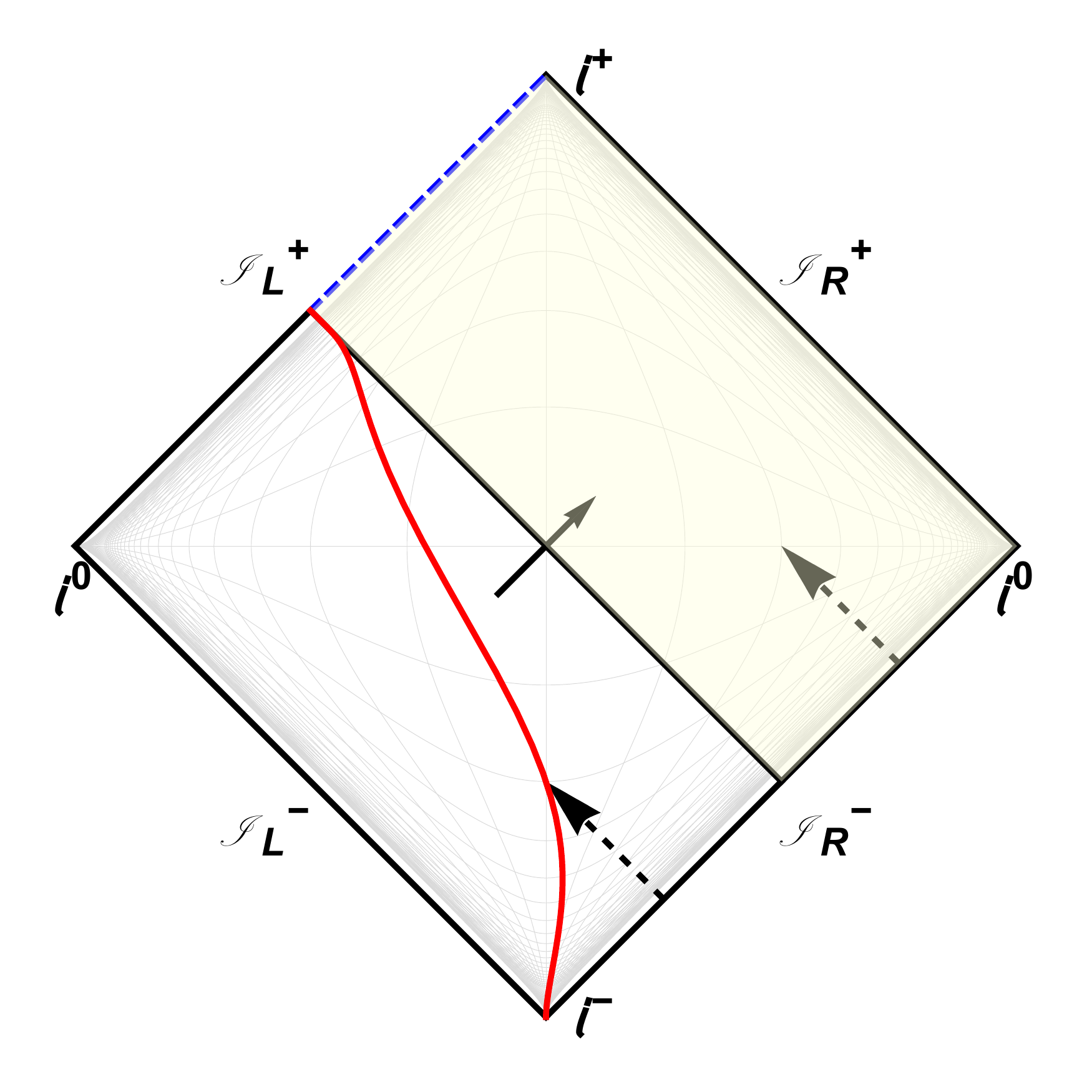} 
\includegraphics[width=2.6in]{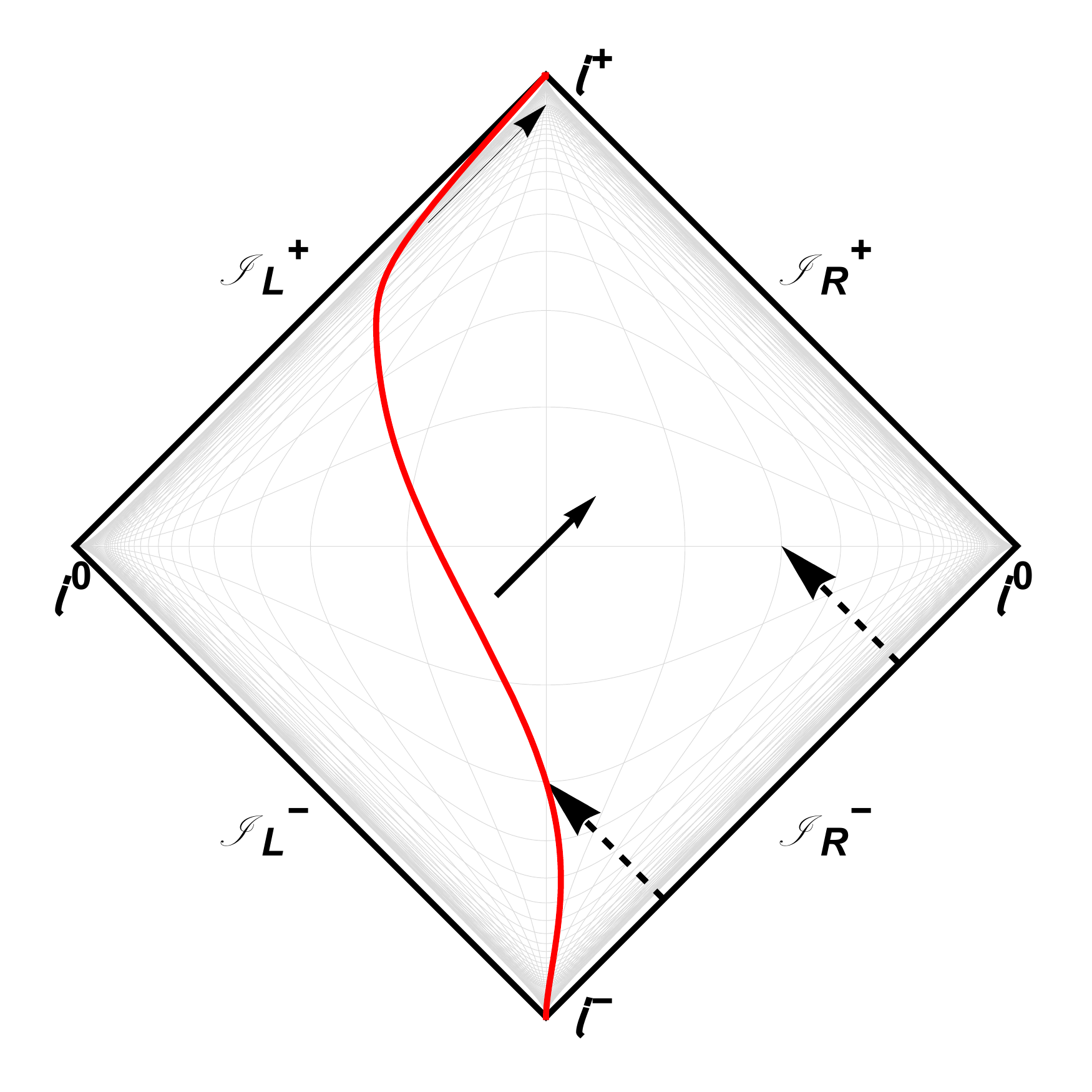} 
\caption{\textbf{Left}: Penrose diagram for the case of the moving mirror (solid red line) which has a one-to-one correspondence to the  usual evaporating black hole with information loss. The solid red line here is Eq.~(\ref{omex}), computed with $\kappa \equiv 1/4M = 1$.  Notice that the mirror goes light-like in the far future. The dashed blue line is the singularity analog where left-movers (opaque yellow region) get `trapped' i.e. they fall into the black hole analog never to return or be observed by $\mathscr{I}_R^+$. \textbf{Right}: Penrose diagram for the case of the moving mirror (solid red line) which corresponds to the remnant end-state with no information loss.  Ultra-late time outgoing reflected modes escape to null-infinity, $\mathscr{I}^+_R$, with a redshift dependent on the final asymptotic speed of the mirror, $1>\xi>0$, where in the diagram, $\xi = 0.9$, for illustration.  In this case, the solid red line, Eq.~(\ref{domex}), is calculated with $\kappa = 1$ and is always time-like, permitting all left-movers to reflect and become right-movers.    } 
\label{fig:penrose_mirror} 
\end{figure}

Prioritizing information preservation requires the existence of an asymptotic comoving observer with the origin.  This is a timelike restriction on the black mirror: its maximum speed must always travel slower than light, even asymptotically\footnote{Interestingly, the speed of light can be timelike approached if the acceleration asymptotes to zero sufficiently fast, see \cite{{Good:2016reflectatlight}}.}.  The black hole origin, $r=0$, which corresponds to the location of the moving mirror, will also be made to be timelike.  Similar to the mirror, the origin, expressed in null coordinates (outside retarded time as a function of inside retarded time) will not asymptotically approach null-future infinity, $\mathscr{I}^+$.   This also means that the situation in the black hole case resolves without $r=0$ attaching to a space-like singularity (to do so would mean that $r=0$ is no longer timelike).  The result of a particular timelike trajectory for the origin is that the usual evaporation process effectively ends with the formation of a remnant, depicted in the Penrose diagram of Fig.~(\ref{fig:penrose}). 
Notice that for the corresponding moving mirror as pictured in the Penrose diagram, Fig.~(\ref{infolossBHC}) or Fig.~(\ref{fig:penrose_mirror}), the worldline never finds itself located at null-future infinity.
To make this quantitative, and to raise the principle of unitarity to the most overarching priority of the model, we modify Eq.~(\ref{omex}) to a new trajectory \cite{Good:2016atu} for the origin of coordinates (one that has an asymptotic drift):
\be t_\xi(x) = - \frac{x}{\xi} - 4 M e^{x/2M\xi}. \label{domex}\ee
Here $\xi$ is the asymptotic coasting speed, $0 < \xi < 1$. This is plotted as mirror in Fig. (\ref{fig:penrose_mirror}). The trajectory in null coordinates changes dramatically from Eq.~(\ref{m1}) to (using the regularity condition result $U\leftrightarrow v$):
\be u_\xi = v - \frac{\xi}{\kappa} \ln\left[\frac{\epsilon}{2}\mathcal{W}\left(\frac{2}{\epsilon} e^{-\frac{2}{\epsilon}\kappa v}\right)\right], \label{m2} \ee
where $\mathcal{W}$ is the Lambert function and we have introduced the definition, $\xi \equiv 1-\epsilon$. 
When $\xi \to 1$, the trajectory Eq.~(\ref{m2}) is equivalent to Eq.~(\ref{m1}), which demonstrates an operative formation of an effective event horizon.  Thus the timelike restriction (by the introduction of a second parameter $\xi$ in addition to mass $M$) in Eq.~(\ref{omex}), resulting in Eq.~(\ref{domex}), permits the genesis of an event horizon \footnote{For all practical purposes.  It is an interesting question whether any type of horizon is formed during gravitational collapse taking into account quantum effects see, e.g. \cite{Mann:2018jcf}. For horizonless models see \cite{Good:2017kjr, Good:2017ddq}.}.

The origin, $r=0$, as far as the quantum field, $\psi$, is concerned, is a perfectly reflecting boundary because $\psi(r\leq 0) = 0$. Eq.~(\ref{m2}) is the timelike world line of the origin and re-tracing our steps from the matching condition Eq.~(\ref{m1}) to Eq.~(\ref{tortoise}), we arrive at:   
\be \label{giant} r^*_\xi \equiv r+ 2 M \xi  \ln \left[\frac{\epsilon}{2} \mathcal{W}\left(\frac{2}{\epsilon} e^{\frac{2}{\epsilon}[\frac{r}{2M}-1]}\right)\right].\ee
In the limit that $\xi \to 1$, Eq.~(\ref{giant}) $\to$ Eq.(\ref{tortoise}), $r^*_{\xi} \to r^*$.  No singularity in Eq.~(\ref{giant}) exists at $r = 2M$.  On the contrary, unlike Eq.~(\ref{tortoise}),  using $\xi \neq 1$ and examining $r=2M$, the function takes on a finite value: $r^*_\xi = 2M \left[1- \xi \mathcal{W}(2/\epsilon)\right]$.  The asymptotic inertial drift, $\xi$, can be close to $\xi \approx 1$ for operative continuity and still be strictly, $\xi < 1$, for information preservation.  At the cost of classical continuity, we have purchased quantum purity.  \\

\section{How Discontinuous?} 
\begin{figure}[h]
\centering 
\includegraphics[width=2.6in]{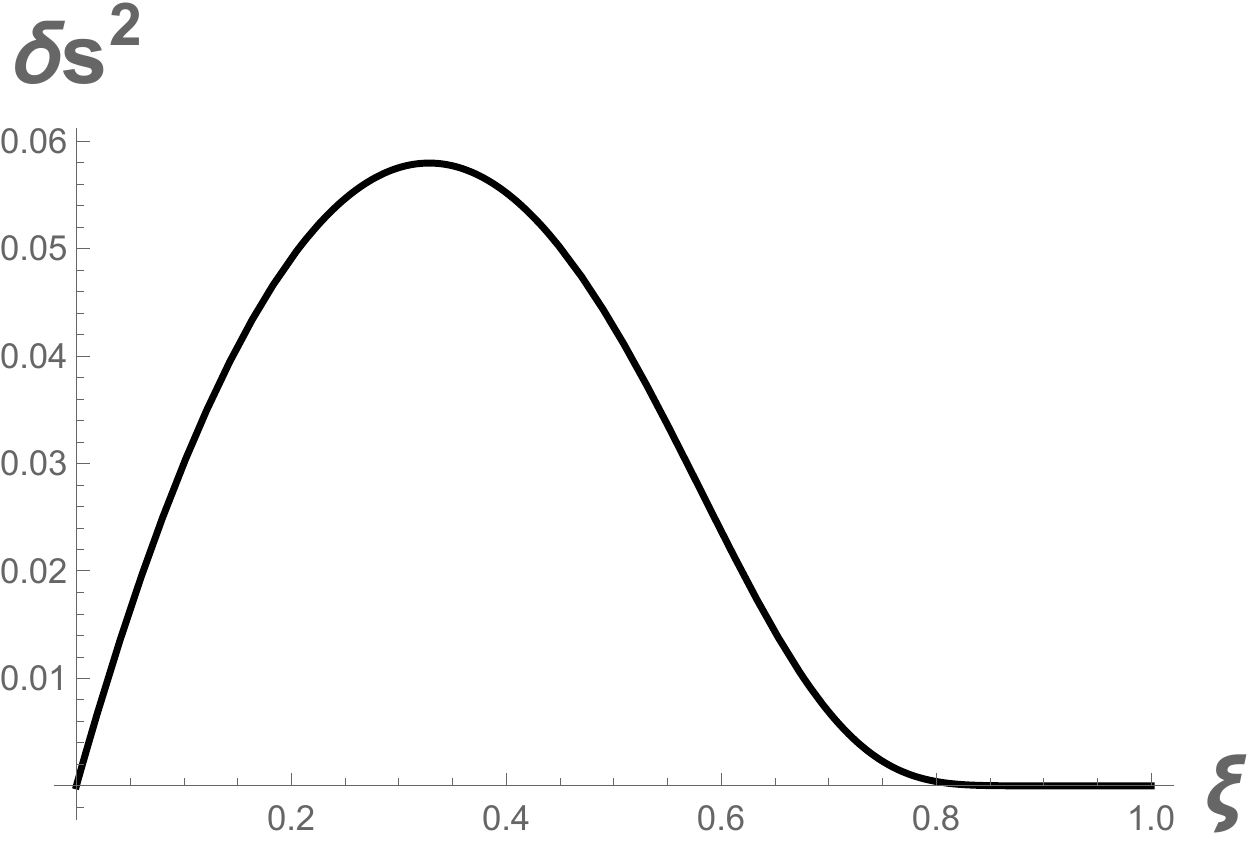} 
\includegraphics[width=2.6in]{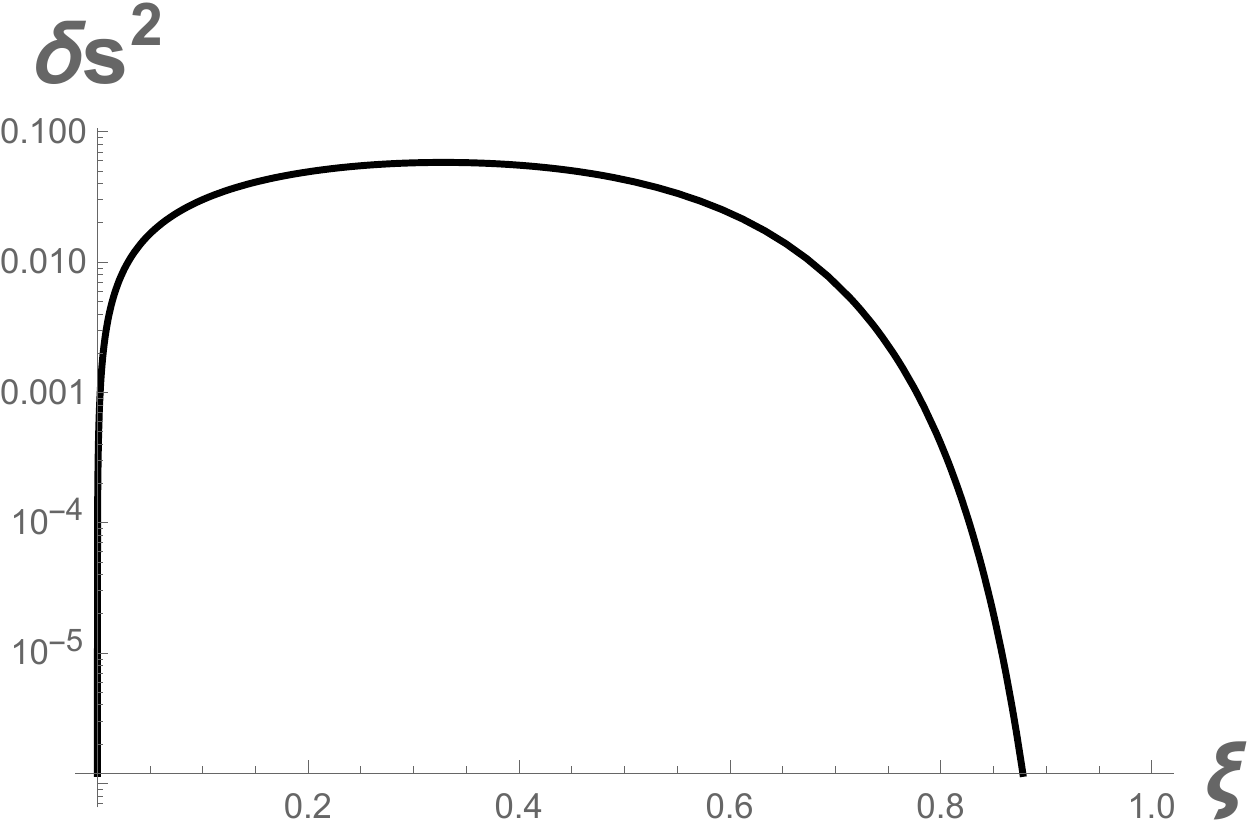} 
\caption{Left: The discontinuity, Eq.~(\ref{zerodiseq}), at late times over the shock wave for ultra-relativistic speed $\xi \gtrapprox 0.9$ is approximately zero. Right: A log plot of Eq.~(\ref{zerodiseq}) to demonstrate the rapid fall off at high speeds. Here $\epsilon \equiv 1 - \xi$, and the final coast speed is greater than zero but less than the speed of light, $0<\xi<1$.} 
\label{fig:zerodis} 
\end{figure}

The discontinuity is located in the Planckian region of Fig. (\ref{fig:penrose}), where one expects quantum geometric effects to play an important role and spacetime discontinuity to possibly occur.  This discontinuity can be characterized at late times by first considering the whole spacetime using the interior coordinates \cite{Wilczek:1993jn}, 
 \begin{equation}\label{in-in}
\d s^2=
\begin{cases}
-\d U\d V,~~~~~~~~~~~~~~~~~~~~~~~~~~~ \text{for}~v\leq v_0, \\
-f(u,v)f^{-1}(u,v_0)\d U\d V,~~\text{for}~v > v_0.
\end{cases}
\end{equation}
The difference at late times, $u\to\infty$, of
\be \delta s^2 = 1-f(u,v)f^{-1}(u,v_0), \ee
is a measure of the discontinuity. The difference, located at the shock wave, $v=v_0$, utilizing Eq.~(\ref{m2}) and its inverse, $v_\xi(u)$, is 
\be \delta  s^2|_{u\to\infty} = \lim_{u\to\infty} \left[1- \frac{d v_\xi}{d u}\frac{d u_\xi}{dv}\right] = 2\frac{ \epsilon -1}{\epsilon -2} \left(1-\frac{1}{W\left(\frac{2}{\epsilon} e^{-2/\epsilon }\right)+1}\right) \label{zerodiseq}\ee
This late-time measure helps clarify how tiny the discontinuity $\delta s^2$ can be when $\xi$ is ultra-relativistic,
\be \delta s^2|_{u\to\infty} \approx 0, \quad \textrm{for}\; \xi \gtrapprox 0.9. \label{zerodiseq}\ee
 as seen in Figure \ref{fig:zerodis}, which also includes a log plot of $\delta s^2$ as a function of $\xi$. 

\section{Restriction on Entropy }

The limit of the von-Neumann entanglement entropy \cite{Chen:2017lum} from the dynamics of Eq.~(\ref{omex}) as a function of lab time $t$ is found from the rapidity \cite{Good:2016atu}, $\eta(t) = \tanh^{-1}\dot{z}(t)$, by using $6S(t) = \eta(t)$:

\be  \bar{S} \equiv \lim_{t\to\infty} S(t) = \lim_{t\to\infty}\frac{1}{6} \textrm{tanh}^{-1}\left[(\mathcal{W}[2e^{-2\kappa t}]+1)^{-1}\right] = \infty. \label{Somex}\ee
This is a consequence of the fact that not all field modes hit the origin and information is lost (see e.g. the $\bar{S} = \infty$ in \cite{GoodS}). In contrast, the asymptotically timelike trajectory of Eq.~(\ref{domex}) has the limit,
\be \bar{S}_\xi \equiv \lim_{t\to\infty} S_{\xi}(t) =  \lim_{t\to\infty}\frac{1}{6} \textrm{tanh}^{-1}\left[\xi (\mathcal{W}[2e^{-2\kappa t}]+1)^{-1}\right]= \frac{1}{6}\textrm{tanh}^{-1}(\xi) \neq \infty. \label{Sdomex}\ee
The final asymptotic entropy, $\bar{S}_\xi$, is the final rapidity (an additive quantity appropriate for an entropy).  This is a result of the fact that all the field modes intersect the timelike worldline origin.  We will see in the next sections that while $\bar{S} = \eta/6$ is finite, it will be extremely large (e.g. relative to usual accelerator rapidities) in order reach effective equilibrium ($\xi \approx 1$). 

In the first case, Eq.~(\ref{Somex}), some left-mover field modes never hit the origin and consequently never become right-movers. These modes are lost forever in the black hole.  However, in the second case, Eq.~(\ref{Sdomex}), an asymptotic approach to time-like future infinity, $i^+$, means that all the left-movers hit the origin and they all become right-movers, preserving information measured by an observer at $\mathscr{I}^+$.  
\section{Restriction on Energy}
Evaluation of the total energy emitted \cite{Good:2016atu}, 
\be E = \int_{-\infty}^{\infty} F(u)\; du, \ee
from the energy flux, $F(u)$, can be done with the Schwarzian derivative \cite{Davies:1976hi,Davies:1977yv},
\be -24 \pi F(u) = \{p,u\} \equiv \frac{p'''}{p'} - \frac{3}{2}\left(\frac{p''}{p'}\right)^2, \ee
of the null-coordinate trajectory, $p(u)$ (the $v$ position of the origin as function of $u$)- in our case, the function $p(u) = v_\xi(u)$, which is the inverse of Eq.~(\ref{m2}).  The result is finite and analytic:
\be E = \frac{\hbar \kappa }{24\pi c}\left(\gamma^2 - \frac{\eta}{\xi}\right), \label{energy1}
\ee
where $\gamma \equiv 1/\sqrt{1-\xi^2}$ is the asymptotic coasting Lorentz factor, $\eta \equiv \tanh^{-1}\xi$ is the asymptotic coasting rapidity, and $\xi<1$ is the asymptotic coasting speed.  [Notice the negative sign which is correct, as opposed to the plus sign misprint in the published version of this note]. 
When $\xi \approx 1$, near the speed of light, then $\gamma^2 \gg \eta/\xi$, and the second term can be neglected:
\be E \sim \kappa \gamma^2. \label{energy} \ee 
The energy diverges as the origin moves asymptotically null, (i.e. $\xi \to 1$ and $\gamma \to \infty$) behaving like the total divergent energy using the inverse of Eq.~(\ref{m1}), $v(u)$, there $\xi = 1$. Consistent with conservation of energy, Eq.~(\ref{energy}) gives the total finite emission when the system has a chance to equilibrate.

\section{Mass of the Remnant}

The energy of the spacetime is $M$ and the evaporation energy is $E$. The difference between them is the mass, $m$, of the remnant:
\be M-E = m, \ee
where for a steady-state system, $\xi \approx 1$, Eq.~(\ref{energy}) holds,
\be M - \frac{\gamma^2}{96 \pi M} = m. \ee
The two parameters of the system, $\gamma$ and $M$, are simply related if the mass of the remnant is so far less than the black hole, $m\ll M$, that it may be neglected:
\be \gamma^2 = 96\pi M^2. \ee
For perspective, a solar mass black hole, $M \sim 10^{38}$ Planck masses, via $\cosh \eta = \gamma$, has final coasting origin rapidity $\eta\sim 91$. The fastest particle ever measured, the OMG particle, had rapidity $\eta \sim 27$. 


\section{Constant Energy Flux and Equilibrium Temperature }
\begin{figure}[h]
\centering 
\includegraphics[width=2.6in]{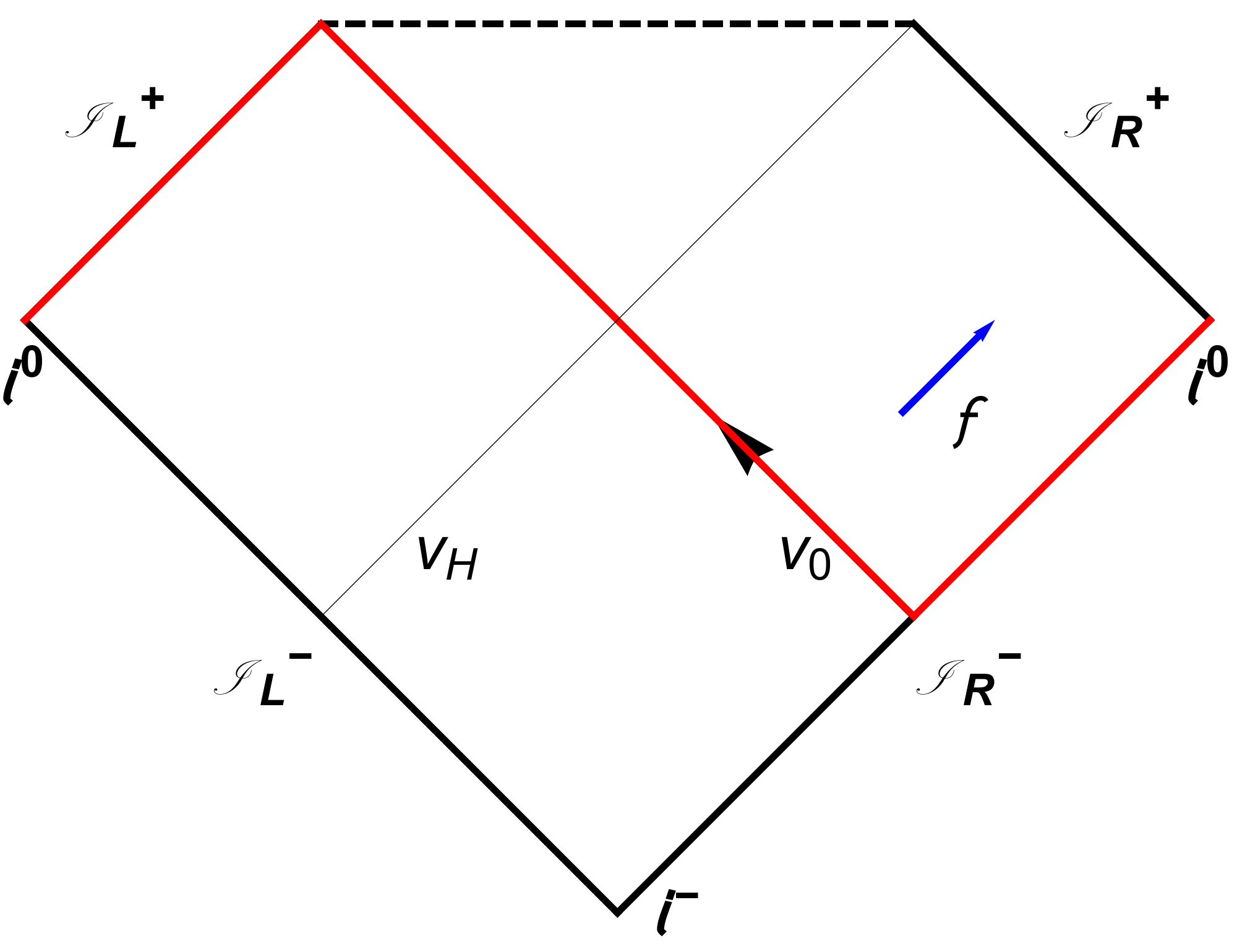} 
\includegraphics[width=2.6in]{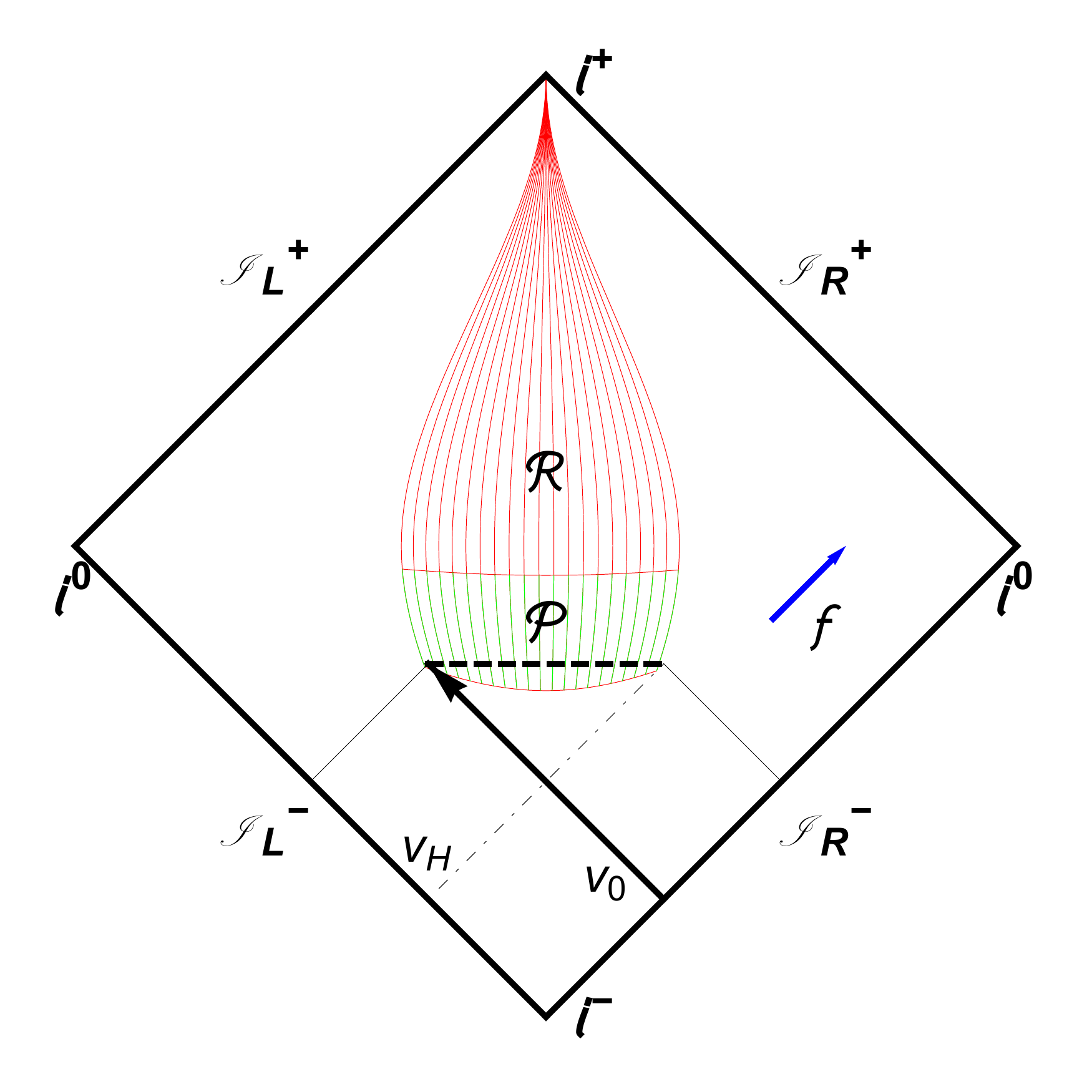} 
\caption{\textbf{Left}: The usual 1+1 Penrose diagram for a Schwarzschild black hole with a singularity and information loss. The solid red line is the Cauchy surface over which integration is done. $v_H$ is the event horizon, and $v_0$ is the shock wave.  The blue arrow, $f$, is Hawking flux.  \textbf{Right}: The 1+1 Penrose diagram for the case of a remnant with infinite life time (solid red lines), region $\mathcal{R}$.  In this situation the remnant end-state allows modes to pass through it (no information loss).  Ultra-late time outgoing modes escape to null-infinity, $\mathscr{I}^+$, with a redshift dependent on the final asymptotic speed of the origin, $1>\xi>0$.  The green lines indicate the Planckian region $\mathcal{P}$ where non-trivial quantum spacetime effects are important. The dotted dashed gray line is the location of the original event horizon.  The dashed black line, inside $\mathcal{P}$, is the location of the original singularity, now hidden. The discontinuity occurs in $\mathcal{P}$ over the shock wave $v_0$ like in the 3+1 case.     } 
\label{fig5} 
\end{figure}

In the 1+1 dimensional black hole case, visually depicted in Fig.~(\ref{fig5}), a temperature, $2\pi T = \kappa$, gives a constant energy flux,
\be F = \frac{\kappa^2}{48\pi} = \frac{\pi}{12} T^2, \ee
which emerges in the form of an extended plateau for assignment of high coast speed, $\xi \approx 1$.  A series expansion of the temperature,
\be T_\xi = \sqrt{\frac{12}{\pi}F^\textrm{max}_\xi}, \ee
as a function of maximum energy flux \cite{Good:2016atu},
\be F^{\textrm{max}}_\xi = \frac{\kappa^2}{48\pi}\left[1 - 3\sqrt[3]{6}\epsilon^{2/3} + O\left(\epsilon\right)\right]\ee
(where the radiation is closest to equilibrium) gives the temperature of the black hole:
\begin{equation}
T_\xi = \frac{\kappa}{2\pi}\left[1-3\left(\frac{3}{4}\right)^{1/3}\epsilon^{2/3}+O\left(\epsilon\right)\right]\label{hot},
\end{equation} to lowest order in $\epsilon \equiv 1-\xi$. The higher order terms correspond to the very small deviation due to sub-light speed drift, $\xi < 1$, which are safely ignored for small $\epsilon \approx 0$. The timelike modification of Eq.~(\ref{omex}) to Eq.~(\ref{domex}) locks in a constant energy flux plateau and effective long-term thermal equilibrium of Eq.~(\ref{hot}) for $\xi \approx 1$. This corroborates the stability of the steady-state (long-time Planckian distributed particle radiation) due to the timelike worldline origin.  

\section{Particle Spectrum}
The beta Bogolubov coefficient in the particle spectrum,
\begin{equation}
\left\langle N_\omega\right\rangle\equiv\langle 0_{\textrm{in}}|N_\omega^{\textrm{out}}|0_{\textrm{in}}\rangle=\int_0^\infty|\beta_{\omega\omega'}|^2d\omega',\label{spectrum}
\end{equation}
can be computed \cite{Good:2016atu} via an integral with input of Eq.~(\ref{domex}).   The result is analytically tractable:
\begin{equation}
\beta_{\xi} = -\frac{\xi\sqrt{\omega\omega'}}{2\pi\kappa \omega_p} \left(\frac{i\kappa}{\omega_p}\right)^{A} \Gamma(A),\label{beta}
\end{equation}
where $A \equiv \frac{i}{2\kappa}[(1+\xi)\omega + (1-\xi)\omega']$ and $\omega_p \equiv \omega+\omega'$. How is this Planckian?  The integrand of Eq.~(\ref{spectrum}) is Planckian using Eq.~(\ref{beta}) with $\xi \approx 1$,
\be |\beta_{\omega\omega'}(\xi\approx 1)|^2 \approx \frac{\omega'}{2\pi \kappa(\omega +\omega')^2}\frac{1}{e^{2\pi \omega / \kappa}-1} \to \frac{1}{2\pi \kappa \omega'}\frac{1}{e^{2\pi \omega / \kappa}-1} \label{betaomex} \ee
where in the second step, we took the high frequency approximation, $\omega'\gg \omega$.  The pre-factor is an continuum artifact removed with wave packets \cite{Hawking:1974sw}. The result Eq.~(\ref{betaomex}) is the usual Planck distribution consistent with Eq.~(\ref{hot}). Further consistency of the result can be shown by using Eq.~(\ref{beta}) to obtain, via a numerical computation, the total energy, 
\be E \equiv \int_0^\infty \hbar  \omega \; \left\langle N_\omega\right\rangle \; d \omega\; \label{totalenergyfrombeta}\ee
and comparing this with the total energy, Eq.~(\ref{energy1}) (both the left and right sides of the origin) from the stress tensor \cite{Good:2016atu, Good:2013lca}.  The results are in agreement.  \\

\section{No Restriction on Total Particle Count: Soft Hair}
It is good to mention a limitation and an extension of this approach.  Assuming unitarity from the start, this model regularizes the total emission energy, the asymptotic entanglement entropy, and the asymptotic rapidity of the origin.  However, the total particle count is still infinite.  That is, the $\xi$-model is still impaired by an infrared divergence in the integrand, $N_\omega$, when computing,
\be N\equiv\int_0^\infty \left\langle N_\omega \right\rangle \; d \omega\; = \int_0^\infty\int_0^\infty|\beta_{\omega\omega'}|^2 d \omega'\; d\omega,\label{totalcount}\ee
which is the total particle count.  Even after the energy has effectively stopped being radiated, an extreme red-shift occurs.  This end-of-time Doppler shift suffered by the modes which continue to pass through the origin marks the existence of the remnant, which is characterized by this divergence, 
\be \textrm{asymptotic drift} : \quad N\to\infty .\ee
The modes reflect off a constant, $\xi > 0$, boundary condition (albeit one with asymptotically zero acceleration). The massless scalar particles carry zero-energy, e.g. \cite{H}.  Generally, this divergence is remedied by abandoning asymptotically coasting trajectories and replacing them with asymptotically static trajectories (the final asymptotic velocity would be zero, $\xi = 0$), which leave no evidence of any remnant (zero Doppler shift), 
\be \textrm{asymptotic static} : \quad N\to \textrm{finite} .\ee
An interesting extension of this model would be a construction of such an asymptotically static dynamic which also has thermal emission and corresponds to Eq.~(\ref{omex}) or Eq.~(\ref{domex}). This line of work is in progress.

\section{Conclusion}
\begin{table}
\centering
\begin{tabular}{|c|c|c|}
	\hline
	Quantity & Continuity \quad $\xi = 1$ & Unitarity \quad $\xi \neq 1$\\
    \hline\hline
    $r^*$ & $r^*\equiv r+2M \ln\left(\frac{r}{2M}-1\right)$ & $r^*_\xi \equiv r+ 2 M \xi  \ln \left[\frac{\epsilon}{2} \mathcal{W}\left(\frac{2}{\epsilon} e^{\frac{2}{\epsilon}[\frac{r}{2M}-1]}\right)\right]$\\
    \hline
    $t(x)$ & $t = - x - 4M e^{x/2M}$ & $t_\xi = - x/\xi - 4 M e^{x/2M\xi}$\\
    \hline
    $u$ & $u=v-4 M \ln|\kappa v|$ & $u_{\xi} = v - 4 M \xi \ln\left[\frac{\epsilon}{2}\mathcal{W}\left(\frac{2}{\epsilon} e^{-\frac{2}{\epsilon}\kappa v}\right)\right]$ \\ 
    \hline
    $\bar{S}$ & $\infty$ & $\bar{S}_\xi = \eta/6$\\
    \hline 
    $E$ & $\infty$ & $E_\xi = \frac{\kappa}{24\pi }\left(\gamma^2- \frac{\eta}{\xi}\right)$\\
    \hline
    $T$ & $T = \kappa/2\pi$ & $T_\xi = \kappa/2\pi + \mathcal{O}(\epsilon),~~\textrm{for} ~~\epsilon\approx 0$\\
    \hline
    $\beta$ & $\beta = -\frac{\sqrt{\omega\omega'}}{2\pi\kappa \omega_p} \left(\frac{i\kappa}{\omega_p}\right)^{\frac{i\omega}{\kappa}} \Gamma(\frac{i\omega}{\kappa})$ & $\beta_{\xi}=-\frac{\xi\sqrt{\omega\omega'}}{2\pi\kappa \omega_p} \left(\frac{i\kappa}{\omega_p}\right)^{A} \Gamma(A)$\\
    \hline
    $N$ & $\infty$ & $\infty$\\
    \hline
   
\end{tabular}
\end{table}

This note postcedes the work of \cite{Good:2016atu,GTC, Myrzakul:2018bhy}. The evaporation model prioritizes unitarity over continuity, and one consequence is that a finite total energy is emitted. Here two analytic answers are found: the beta Bogolubov coefficient and the finite total energy.  The dynamics are described by the asymptotically timelike worldine, Eq.~(\ref{m2}), and the two-parameter Regge-Wheeler coordinate, Eq.~(\ref{giant}).  

Uncompromising continuity across the shock wave in the metric results in information loss.  At the Planck scale, at least, continuity is less certain.  High precision in effective continuity is permitted with very fast sub-light drifting speeds. This relaxation results in preserved information, finite energy and radiated Planckian distributed particles with constant energy flux at equilibrium temperature.  

\section*{Acknowledgments}
M.R.R.G. thanks Yen Chin Ong, Aizhan Myrzakul, and Khalykbek Yelshibekov. M.R.R.G.
also thanks Daniele Malafarina for clarifying points about the Penrose diagrams. The Ministry of Education and
Science of the Republic of Kazakhstan are also acknowledged.


\end{document}